\documentclass[pdflatex,sn-nature]{sn-jnl}

\usepackage{graphicx}%
\usepackage{multirow}%
\usepackage{amsmath,amssymb,amsfonts}%
\usepackage{amsthm}%
\usepackage{amsmath}
\usepackage{mathrsfs}%
\usepackage[title]{appendix}%
\usepackage{xcolor}%
\usepackage{textcomp}%
\usepackage{manyfoot}%
\usepackage{hyperref}
\usepackage{booktabs}%
\usepackage{algorithm}%
\usepackage{algorithmicx}%
\usepackage{algpseudocode}%
\usepackage{listings}%
\usepackage{longtable}

\theoremstyle{thmstyleone}%
\theoremstyle{thmstyletwo}%

\theoremstyle{thmstylethree}%

\providecommand\aj{Astron.\ J.} 
\providecommand\araa{ARA\&A} 
\providecommand\aap{Astron.\ Astrophys.} 
\providecommand\apj{Astrophys.\ J.} 
\providecommand\apjl{Astrophys.\ J.} 
\providecommand\icarus{Icarus} 
\providecommand\pasp{PASP} 
\providecommand\mnras{Mon.\ Not.\ R.\ Astron.\ Soc.} 
\providecommand\psj{Planet.\ Sci.\ J.} 

\begin{document}

\title[3I carbon and nitrogen isotopes]{High nitrogen and carbon isotopic ratios in the interstellar comet 3I/ATLAS}


\author*[1]{Opitom, C.$^\dagger$}\email{copitom@ed.ac.uk}

\author[2]{Manfroid, J.$^\dagger$}

\author[2]{Hutsemékers, D.$^\dagger$}

\author[2]{Jehin, E.}
\author[3]{Knight, M.~M.}
\author[2]{Aravind, K.}
\author[4]{Ferellec, L.}
\author[5]{Bodewits, D.}
\author[6]{Guzmán, V. V. }
\author[7,8]{Cordiner, M.}
\author[9]{Dorsey, R. C.}
\author[10]{La Forgia, F.}
\author[11]{Lippi, M.}
\author[1]{Murphy, B. P.}
\author[1]{Snodgrass, C.}
\author[12]{Bannister, M.}

\footnotetext{$^\dagger$ These authors contributed equally to this work.}
\affil*[1]{Institute for Astronomy, University of Edinburgh, Royal Observatory, Edinburgh EH9 3HJ, UK}

\affil[2]{STAR Institute, University of Liège, Allée du 6 août, 19, 4000 Liège (Sart-Tilman), Belgium}

\affil[3]{Volgenau Department of Physics, United States Naval Academy, 572C Holloway Road, Annapolis, MD 21402, USA}

\affil[4]{Faculty of Science and Engineering, Northumbria University, Newcastle NE1 8ST, UK}

\affil[5]{Physics Department, Edmund C. Leach Science Center, Auburn University, Auburn, AL 36849, USA. }

\affil[6]{Instituto de Astrofísica, Pontificia Universidad Católica de Chile,
Av. Vicu na Mackenna 4860, 7820436 Macul, Santiago, Chile}

\affil[7]{Department of Physics, Catholic University of America, Washington, DC 20064, USA.}

\affil[8]{Astrochemistry Laboratory, NASA Goddard Space Flight Center, 8800 Greenbelt Road, Greenbelt, MD 20771, USA.}

\affil[9]{Department of Physics, P.O. Box 64, 00014 University of Helsinki, Finland}

\affil[10]{Department of Physics and Astronomy, University of Padova, vicolo Osservatorio 3, 35020 Padova, Italy.}

\affil[11]{INAF - Osservatorio Astrofisico di Arcetri, Largo Enrico Fermi 5, 50125 Firenze, Italy}

\affil[12]{School of Physical and Chemical Sciences – Te Kura Matū, University of Canterbury, Private Bag 4800, Christchurch, 8140, Aotearoa New Zealand.}


\abstract{
Interstellar objects provide a unique opportunity to further our understanding of the planetary formation process by studying in detail material formed around another star. 
Their ices contain precious clues about the environment and conditions prevailing in their home system.  As fractionation processes can be sensitive to the temperature and radiation environment, isotopic ratios are powerful tracers of the origin and evolution of different species. While isotopic ratios have been measured in solar system comets, previously detected interstellar objects have been too faint to measure isotopic ratios. 
Here we report the measurement of two ratios in 3I/ATLAS from observations of the CN molecule: $^{12}$C/$^{13}$C and $^{14}$N/$^{15}$N. We report $^{12}$C/$^{13}$C~=~$147^{+87}_{-40}$ and $^{14}$N/$^{15}$N~=~$343^{+454}_{-124}$. The $^{14}$N/$^{15}$N is higher than the value of $\sim$~150 usually measured for solar system comets, close to the values measured in the interstellar medium, pre-stellar phases or the outside of protoplanetary discs. The $^{12}$C/$^{13}$C is marginally higher than the values usually measured for solar system comets and in the interstellar medium.
These measurements could indicate an origin of 3I in the outer disc around an older low-metallicity star.}




\maketitle

Interstellar objects, formed in  planetary systems beyond our own and now passing through the solar system, provide a rare opportunity to study material formed in another proto-planetary disc potentially having very different physical and chemical conditions. When such objects become active and sublimate, the released gases can be studied spectroscopically, allowing us to directly probe their volatile composition and isotopic ratios. While modern instrumentation allows us to study protoplanetary discs remotely, their large distances from us constrains the level of detail that can be obtained. By contrast, studying interstellar objects at a relatively close distance while they are passing through the solar system provides a unique window into the conditions prevailing in these discs where planetesimals and planets are forming. The composition of their ices, and especially the relative abundances of isotopes, constitute invaluable probes into their formation conditions. 

The first two interstellar objects, 1I/`Oumuamua, discovered in 2017 \citep{Williams2017} and 2I/Borisov discovered in 2019\footnote{\textit{Minor Planet Electronic Circular} 2023-U162} (hereafter 1I and 2I, respectively), were studied intensively by astronomers all around the world. However, no gas was detected around 1I and the constraints put on the composition of 2I were limited by its relatively low brightness. 

3I/ATLAS (hereafter 3I) was discovered in July 2025 by the ATLAS survey at about 5 au from the Sun and already active \citep{Denneau2025,Jewitt2025a,Alarcon2025,Seligman2025}. 3I was discovered several months before its perihelion which occurred on 29th October 2025 and was significantly brighter than previously discovered interstellar objects. Early observations indicated that 3I may have a composition different from that of most solar system comets. JWST observations at 3.3~au pre-perihelion revealed a coma extremely rich in CO$_2$ and to a lesser degree CO, relative to water \citep{Cordiner2025}. High spectral-resolution optical observations unveiled a very high nickel abundance and, once the comet approached perihelion, the presence of iron in the coma, with a Ni~\textsc{i}/Fe~\textsc{i} abundance ratio initially exceeding that observed in solar system comets \citep{Rahatgaonkar2025,Hutsemekers2025}. In addition, \cite{Roth2025} reported an exceptionally high CH$_3$OH/HCN ratio, higher than measured in all but one solar system comet. 

Together, these compositional signatures suggest that 3I formed under conditions markedly different from those that prevailed in the Solar System. Isotopic ratios are often used to trace the origin and evolution of different species. As fractionation processes can be sensitive to the temperature and radiation environment, isotopic ratios allow us to follow the chemical evolution of material from the pre-stellar stage, through the protostellar and protoplanetary disc stages, and into fully formed planets and planetesimals. As such, isotopic ratios provide a very sensitive probe of the formation conditions. Within the solar system, isotopic ratios of C, N, O, and H  have been measured in a wide range of bodies, revealing substantial variations measured for D/H or $^{14}$N/$^{15}$N, e.g. \cite{Nomura2023}.

Nitrogen isotope ratios are particularly diagnostic as they can be significantly modified by disc chemistry, especially through selective photo-dissociation which is dependent on the radiative environment within the disc. Early models proposed that nitrogen isotope fractionation was primarily driven by isotope exchange reactions \citep{Wirstrom2012}, but recent investigations have shown that these reactions are much less efficient than originally thought \citep{Wirstrom2018}. An alternative mechanism involves the selective photo-dissociation of N$_{2}$ and the subsequent formation of species enriched in $^{15}$N, including HCN \cite{Nomura2023}. At small distances from the star, high N$_{2}$ abundances and strong UV radiation lead to N$_{2}$ self-shielding. This effect favours the photodissociation of N$^{15}$N, lowering $^{14}$N/$^{15}$N ratios in the inner disc. This is consistent with the trend of increase in $^{14}$N/$^{15}$N ratios with the stellar distance measured from observations of HCN in protoplanetary discs \cite{Hily-Blant2019,Rampinelli2025}.

The ratio between the two stable nitrogen isotopes ($^{14}$N/$^{15}$N) has been measured in a range of solar system comets, using different molecular tracers.  Observations of CN revealed an enrichment in heavy nitrogen by a factor of three  relative to proto-solar value \cite{Manfroid2009}. Similar enrichments were subsequently found in HCN and later in NH$_2$, see the review in \cite{Biver2024}. In situ measurements by the ROSINA mass spectrometer on board Rosetta determined $^{14}$N/$^{15}$N~$\sim 130$ in  molecular nitrogen in comet 67P/Churyumov-Gerasimenko \citep{Altwegg2019}. This value is consistent with what is measured from observations of CN, HCN and NH$_2$ for other comets, thus ruling out the hypothesis of the existence of two distinct nitrogen reservoirs in comets. Overall, the measured values are remarkably uniform across solar system comets of different dynamical origins. One notable exception is the lower heavy nitrogen enrichment in split comet 73P/Schwassmann-Wachmann 3 (hereafter 73P) \citep{Manfroid2009}.  More recently, \cite{Cordiner2024} reported a $^{14}$N/$^{15}$N ratio of $68\pm27$ in comet 46P/Wirtanen from HCN observations, suggesting a larger diversity of nitrogen fractionation among solar system comets than originally thought. However, independent measurements of CN in the same comet yielded a value of $150\pm30$ \cite{Moulane2023}, consistent with what is measured for most solar system comets. The origin of the $^{15}$N enrichment in solar system comets remains debated but is most likely the result of the above mentioned isotope-selective photo-dissociation of N$_{2}$ in the protoplanetary disc. 

Fractionation of carbon isotopes is thought to occur through selective photodissociation of CO and exothermic isotope exchange reactions. Models predict that different levels of fractionation through isotope exchange reaction are expected between molecules formed from CO (like CO$_2$) and molecules formed from C$^+$ (like HCN and CN), with molecules formed from CO richer in $^{13}$C \citep{Langer1984}. However, this has been challenged by measurements of $^{12}$C/$^{13}$C in dense clouds, with values either similar to the Interstellar Medium (ISM) or enriched in $^{13}$C \citep{Nomura2023}. 
Measurements of $^{12}$C/$^{13}$C isotopes in solar system comets were originally made from observations of C$_2$ and CN, with values around 90, consistent with solar and terrestrial values \cite{Manfroid2009,Biver2024}. In-situ measurements in CO, CO$_2$, CH$_4$, C$_2$H$_6$ and CH$_3$OH by the ROSINA mass spectrometer onboard Rosetta at comet 67P revealed similar values in the range 84-91, with only H$_2$CO showing a lower value of 40 \citep{Rubin2017,Hassig2017,Altwegg2020}.

\begin{figure*}
        \centering
        \resizebox{0.95\hsize}{!}{\includegraphics{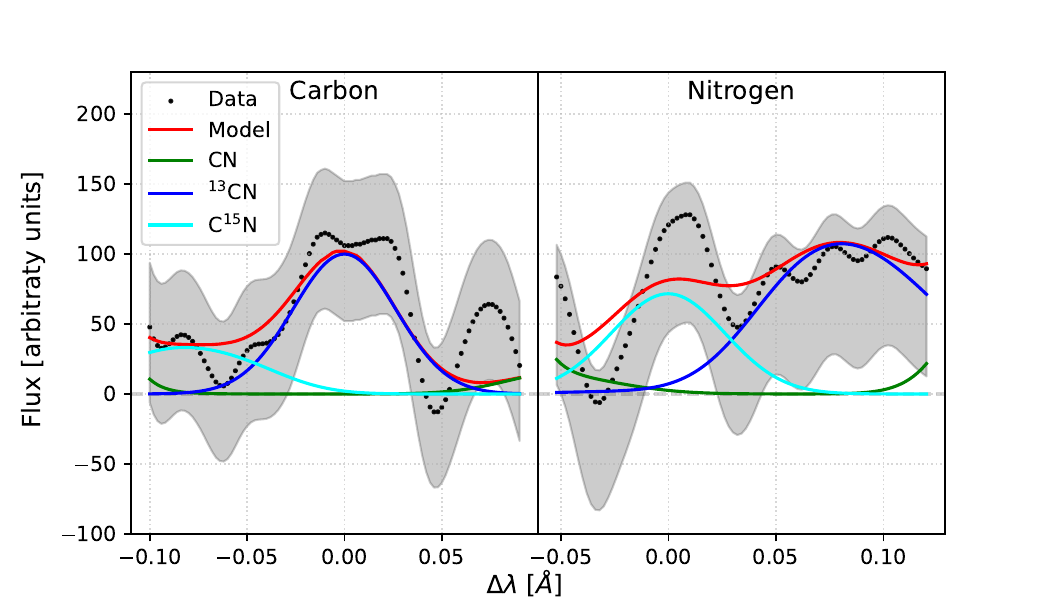}}
        \caption{\textbf{Co-added observed spectra of 3I with best-fit model} Observed and modelled $^{13}$CN-centred (left) and C$^{15}$N-centred (right) line profiles. The shaded region represents the uncertainty of the profile resulting from the combination of the different lines, estimated from the MCMC fitting.   
        \label{Fit_results}
        }%
    \end{figure*}

We performed observations of 3I with the UV-Visual Echelle Spectrograph (UVES) on the Very Large Telescope between 6 and 26 December 2025, from which we measure isotopic ratios for $^{14}$N/$^{15}$N and $^{12}$C/$^{13}$C in CN. Details about the observations and data analysis are presented in the Methods section \ref{methods}. We measure $^{12}$C/$^{13}$C~=~$147^{+87}_{-40}$ and $^{14}$N/$^{15}$N~=~$343^{+454}_{-124}$. The spectra co-added for all dates with lines combined and subtracted from CN are presented in Fig. \ref{Fit_results} for both $^{13}$CN and and C$^{15}$N. The uncertainties given are those obtained through the jackknife resampling, as explained in the methods section, but are of the same order of magnitude as the 3-sigma uncertainties from the Markov Chain Monte Carlo (MCMC) fit.

\begin{figure}
        \centering
        \resizebox{0.95\hsize}{!}{\includegraphics{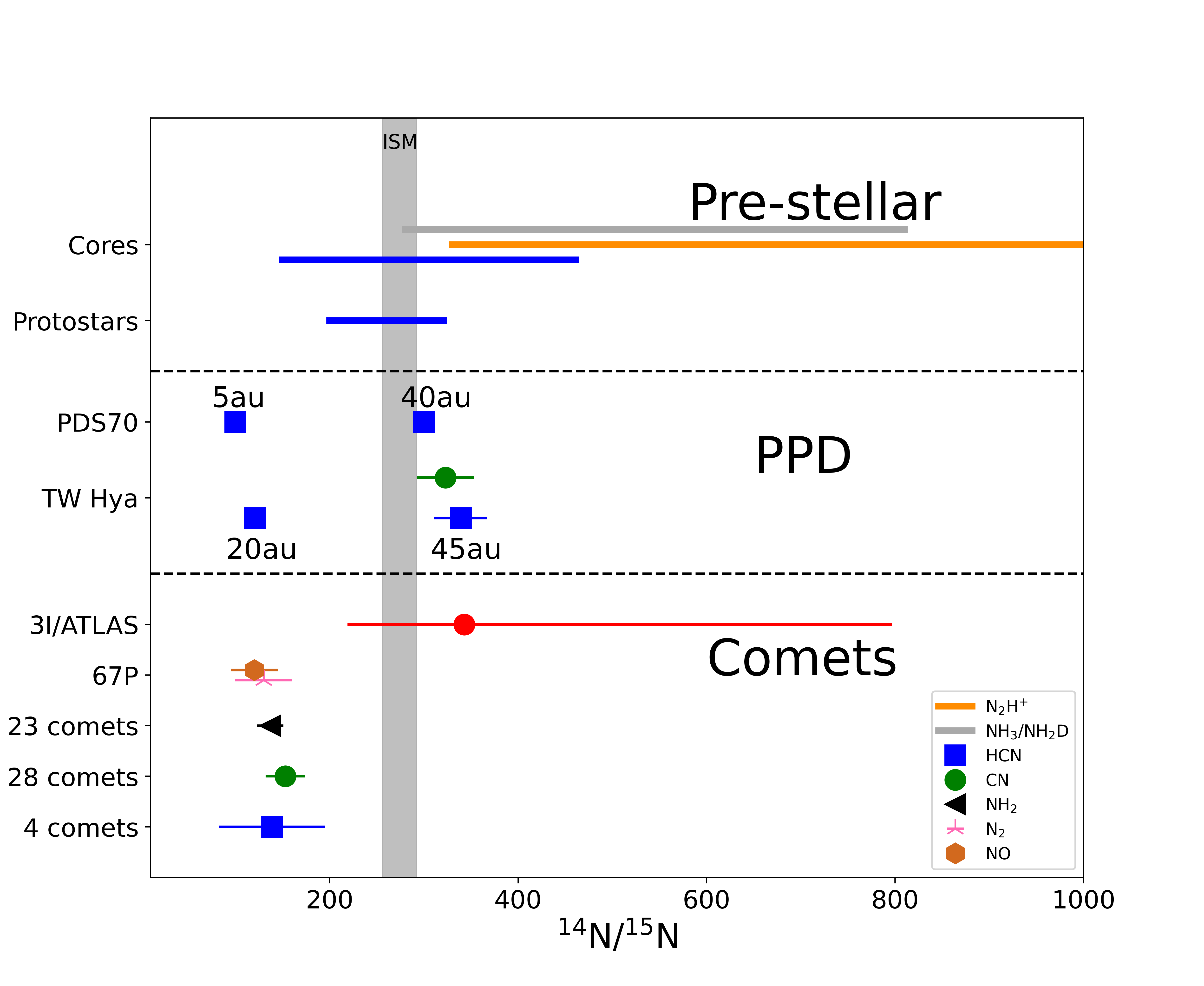}}
        \caption{\textbf{Comparison of $^{14}$N/$^{15}$N in different types of objects} $^{14}$N/$^{15}$N isotopic ratios in solar system comets measured from HCN, CN, and NH$_2$ are at the bottom, with different molecules represented by different symbols (values and references are listed in Table~\ref{tab:n_ratios}). 3I is the red circle, with a $^{14}$N/$^{15}$N ratio measured in CN. Values for protoplanetary discs, pre-stellar cores, and protostars follow the same convention. Values from protoplanetary discs (PPD) come from \cite{Hily-Blant2019,Rampinelli2025}. Prestellar $^{14}$N/$^{15}$N values come from \cite{Jensen2024} for HCN and from  \cite{Nomura2023} for the other molecules.
        \label{Nisotopes}
        }%
    \end{figure}

The $^{14}$N/$^{15}$N ratio in 3I is compared to values measured in solar system comets and other environments in Fig.~\ref{Nisotopes}. This is the first measurement of the nitrogen isotopic ratio in an interstellar comet. The $^{14}$N/$^{15}$N ratio measured for 3I is significantly larger than the value of 150 that is the average measured for solar system comets. It is also higher but consistent within the uncertainties to the ISM value of $274\pm18$ \citep{Ritchey2015} and close to the solar value of $458.7\pm4.2$ \citep{Marty2011}. It is consistent with values measured in HCN in pre-stellar cores and proto-stars, which vary between $\sim$ 150 and 450 depending on the target \citep{Jensen2024}, as well as with even higher values measured for other species \citep{Nomura2023}. The $^{14}$N/$^{15}$N abundance ratio has been measured in star formation regions across the Milky Way, with values ranging from $\sim$~200 to $\sim$~800. \cite{Colzi2022} report a trend of slightly increasing $^{14}$N/$^{15}$N ratios until about 11~kpc from the galactic centre, with a decrease after this distance. Given the spread of values even at similar galactocentric distances and large uncertainties, it is impossible to pinpoint an origin of 3I in a specific region of the galaxy or around a specific type of star from the value of the nitrogen isotopic ratios. 

The $^{14}$N/$^{15}$N ratio has been measured in several proto-planetary discs. Disc-integrated measurements of HCN initially revealed a higher abundance of heavy nitrogen compared to pre-stellar cores \cite{Nomura2023}, with values consistent with solar system comets and about a factor two lower than what we measure for 3I. \cite{Guzman2017} provided the first tentative evidence for increasing $^{14}$N/$^{15}$N ratios with distance from the host star in protoplanetary discs. This was further confirmed by \cite{Hily-Blant2019} from observations of the T~Tauri star TW Hya. They measure a disc-average $^{14}$N/$^{15}$N ratio of 223$\pm$21, with values of 121$\pm$11 at 20 au increasing to 339$\pm$28 at 45 au. More recently, \cite{Rampinelli2025}, reported a ratio increasing from about 100 at 5 au for PDS 70, peaking at more than 300 around 40 au and then decreasing at larger distances from the star. These trends have been attributed to shielding in the inner part of proto-planetary discs, making the selective photo-dissociation of N$_{2}$ less efficient and preventing enrichment in $^{15}$N. The $^{14}$N/$^{15}$N we measure could indicate that 3I was formed in a part of the disc where the selective photo-dissociation of N$_2$ is not very efficient. This could indicate a formation at a relatively large distance from the star or more generally in an environment where shielding prevents selective photo-dissociation of N$_2$. An origin at a relatively large distance from the star would also be consistent with the high CO and CO$_2$ abundances reported by \cite{Cordiner2025}.

\begin{figure}
        \centering
        \resizebox{0.95\hsize}{!}{\includegraphics{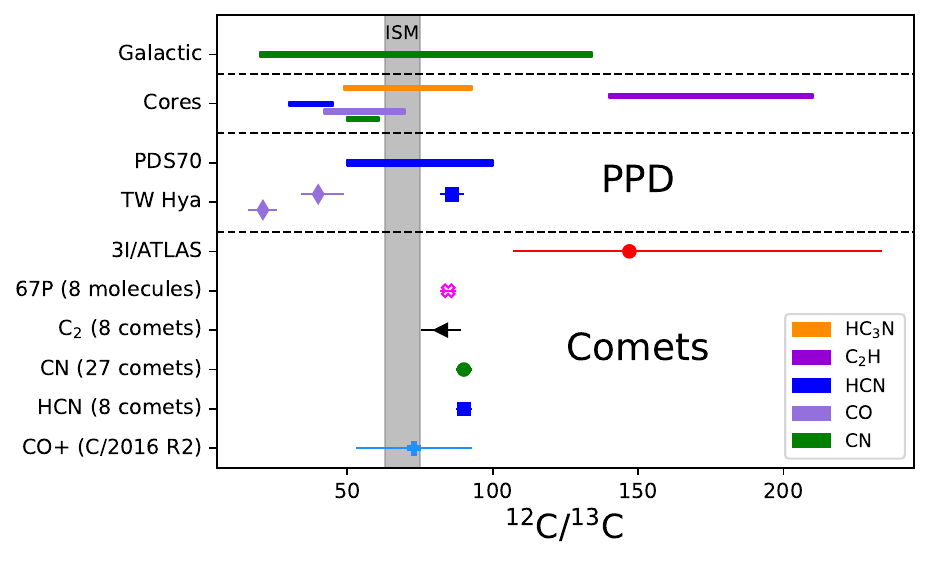}}
        \caption{\textbf{Comparison of $^{12}$C/$^{13}$C in different types of objects} $^{12}$C/$^{13}$C isotopic ratios in solar system comets were measured from various molecules, with different molecules represented by different symbols. Plotted values are the weighted average and uncertainty of all published value listed in Table~\ref{tab:c_ratios}; in some cases multiple measurements were included for a single comet. The 67P point represents the weighted average of nine measurements of eight molecules obtained in situ during the Rosetta mission. 3I is the red circle. Values for protoplanetary discs \citep{Zhang2017, Hily-Blant2019, Yoshida2022, Rampinelli2025}, pre-stellar cores \citep{Nomura2023}, galactic \citep{Milam2005}, and ISM \citep{Wilson1999} follow the same color convention and are identified in the legend. 
        \label{Cisotopes}
        }%
    \end{figure}

The $^{12}$C/$^{13}$C isotope ratio in 3I is higher than most of the values measured in CN for solar system comets, which average around 90 and range from 65 to ${\sim}100$ \citep{Biver2024}, but marginally consistent within the uncertainties. This is illustrated in Fig.~\ref{Cisotopes}. The value we measure is higher than the Interstellar Medium value of $69\pm6$ \cite{Wilson1999}. $^{12}$C/$^{13}$C was constrained for the first time in an interstellar object by \cite{Cordiner2025}, who set a lower limit of $^{12}$C/$^{13}$C $>63$ in 3I from detection of CO$_2$ and $^{13}$CO$_2$ using NIRSPEC on the JWST. More recently, \cite{Cordiner2026} report a $^{12}$C/$^{13}$C ratio in CO$_2$ and CO in the range 129--196 from JWST observations. This is consistent with our measurement of the same ratio in CN.  

Modelling by \cite{Hopkins2025} and work by \cite{Taylor2025} suggests that 3I originated around an old low metallicity star. Studies of the $^{12}$C/$^{13}$C ratio across the galaxy reveal a gradient of the isotopic ratio, with lower values (20-25) towards the centre of the Milky way and larger (up to $\sim$130) ratios at large galactocentric distances \citep{Yan2023,Milam2005}. This is attributed to the degree of stellar nucleosynthesis in different regions of the galaxy; as the galaxy becomes more chemically complex, more $^{12}$C is produced relative to $^{13}$C, for example through the CNO cycle in asymptotic giant branch stars \citep{Timmes1995}. Chemical evolution models presented by \cite{Kobayashi2011} confirm that high $^{12}$C/$^{13}$C ratios are expected around low metallicity stars. The relatively low abundance of $^{13}$C we measure could support an origin of 3I around a low-metallicity star.

Processes occurring during the stellar and planetary formation stages could result in even higher $^{12}$C/$^{13}$C ratios. The $^{12}$C/$^{13}$C ratio in 3I is higher than that reported for protoplanetary discs; \cite{Hily-Blant2019} measure $86\pm4$ from HCN in the TY Hya protoplanetary disc and \cite{Rampinelli2025} report values $<100$ for PDS70 at stellar distances smaller than 100 au. Contrary to the nitrogen ratio, there is no strong indication of variation of the carbon isotopic ratio within discs, except potentially for PDS70 at large distances from the star \citep{Rampinelli2025}. Models predict that the dominant mechanism for carbon isotope fractionation in discs is the $$\text{CO} + \text{}^{13}\text{C}^+ \rightleftharpoons \text{}^{13}\text{CO} + \text{C}^+ + 35 \,\text{K}$$ exothermic exchange reaction. In that scenario, molecules formed from C$^+$ (like CN) are $^{13}$C-poor, while molecules formed from CO are $^{13}$C-rich. As demonstrated in \cite{Nomura2023}, scenarios where the absorption is low, leading to sufficient far ultra-violet radiation, can result in high $^{12}$C/$^{13}$C in HCN. This would be consistent with 3I forming in the outer disc, where photons can scatter and reach deeper layers. Recent work by \cite{Lee2024} also investigated the effect of the carbon to oxygen elemental ratio in proto-planetary discs. Their model indicates that an abundance ratio of C/O $\sim 1-1.5 $ (higher than the C/O in solar twins $\sim~0.5$ or the Sun $\sim~0.55$ \citep{Bedell2018,Nissen2015}) can result in H$^{12}$CN/H$^{13}$CN ratios larger than 100. This could be consistent with the large abundance of CO and CO$_2$ relative to water reported by \cite{Cordiner2025}. These scenarios would likely result in $^{12}$C/$^{13}$C ratios in molecules like CO and CO$_2$ lower than what we measure for CN. However, this is not consistent with measurement of the $^{12}$C/$^{13}$C ratio in CO and CO$_2$ by JWST.

Work by \cite{Maggiolo2026} indicated that the long residence of 3I in interstellar space likely resulted in galactic cosmic ray processing of the surface layers of 3I, which could affect the composition of the gas observed, particularly pre-perihelion. Our observations were all performed post-perihelion, at a time when the comet had been very active for several months. This resulted in outgassing from deeper layers likely unaffected by processing at the time of our observations. 

In summary, using observations of the CN molecule with the UVES spectrograph at the VLT, we measure a high $^{14}$N/$^{15}$N isotopic ratio in the interstellar comet 3I, a factor of two higher than what is measured in solar system comets but consistent with measurements of the same ratio in CN or HCN in pre-stellar cores, protostars, and at larger stellar distances in protoplanetary discs. We measure a $^{12}$C/$^{13}$C ratio higher than but marginally consistent with solar system comet values. This is consistent with values expected for older, low metallicity stars.

\section{Methods}\label{methods}

\subsection{Observations}
Observations were performed with the UV-Visual Echelle Spectrograph (UVES\footnote{UVES User Manual, VLT-MAN-ESO-13200-1825, \\ https://www.eso.org/sci/facilities/paranal/instruments/instruments.html}) at the European Southern Observatory (ESO) Very Large Telescope (VLT) Unit Telescope 2 (UT2), under programme 115.28NL.003. Observations were obtained as soon as possible after perihelion, over a period ranging from 6 to 26 December 2025. While observations with UVES as part of this programme span a longer period of time, we selected observations with the highest signal to noise ratio for this work. We used two different settings to cover the entire optical range: 348+580 (dichroic 1) covering the ranges 3050-3888 \AA$ $ in the blue and 4760-6840 \AA$ $ in the red, and 437+860 (dichroic 2) covering the range 3740-4990 \AA $ $ in the blue and 6600-10600 \AA$ $ in the red. The slit width was set to 0.6$"$, providing a resolving power of $\sim$60,000. In this manuscript, we will focus on observations with the blue arm of each setting as they contain the CN B$^{2}\Sigma^{+}$-X$^{2}\Sigma^{+}$ (0,0) emission band which we are interested in (around 3880 \AA). The observational circumstances of the spectra used in this work are presented in Table~\ref{obs}.

\begin{table}[ht!]
\caption{\label{obs}}
\centering
\begin{tabular}{lcccccc}
\hline\hline
Date (UT) & Set-up &$r_h$ & $\dot{r_h}$ & $\Delta$ & $\dot{\Delta}$ & Exp. Time\\
DD-MM-YYYY HH:MM-HH:MM &   & au & km/s & au & km/s & \#~$\times$ s \\
\hline
06-12-2025 07:25-08:25 & 348  &   1.93 & 44.5 & 1.86 & -15.5 & 3$\times$1000 \\
10-12-2025 07:10-08:35 & 437  &   2.04 & 46.5 & 1.83 & -11.6 & 3$\times$1600 \\
15-12-2025 07:05-07:56 & 348  &   2.18 & 48.6 & 1.80 & -5.8 & 3$\times$1300 \\
21-12-2025 05:40-07:35 & 348  &   2.35 & 50.6 & 1.80 &  2.8 & 2$\times$3400 \\
21-12-2025 07:40-08:35 & 437  &   2.35 & 50.6 & 1.80 & 3.1 & 1$\times$3400 \\
26-12-2025 07:05-07:55 & 348  &   2.50 & 51.9 & 1.82 & 11.3 & 1$\times$3200 \\
\hline
\end{tabular}
\end{table}

\subsection{Data reduction}
Before processing with the ESO pipeline, cosmic ray hits were removed from the raw frames using the ``lacosmic'' package \citep{vanDokkum2001}. The data were then reduced using the ESO UVES pipeline, resulting in flux and wavelength-calibrated 2D spectra. The spectrum extraction (over the full slit length) and subtraction of the continuum were performed using custom python scripts following the procedure described in \cite{Manfroid2009}, resulting in wavelength and flux-calibrated 1D spectra. The wavelength calibration was refined using CN and solar lines, yielding an accuracy better than 0.1 km/s. Particular care was taken  to remove solar light scattered by cometary dust and by the Earth’s atmosphere (twilight and moonlight).

\subsection{Modelling and data fitting}
Fluorescence models of the three isotopologues ($^{12}$C$^{14}$N, $^{13}$C$^{14}$N, $^{12}$C$^{15}$N) were computed for each spectrum using the model described in \cite{Manfroid2009}. The instrumental line profile was adjusted to the observations through a convolution kernel
computed with the IRAF $psfmatch$ procedure. The observed spectra and the models were then optimally combined according to their signal‑to‑noise ratios.
Isotopic abundances were derived from the combined profiles of selected lines for each isotopologue. Combinations of lines were done separately to produce a final observed spectrum centred for $^{13}$CN and C$^{15}$N respectively. Isotopic abundances were derived by simultaneously fitting a combination of the CN, $^{13}$CN and C$^{15}$N for both the $^{13}$CN-centred and C$^{15}$N-centred profiles. The background subtraction described was not perfect, so different fitting methods were tested to account for this, which all gave similar results. Isotopic abundances presented here were derived by minimizing the standard deviation of the residuals between the model and the observations for the carbon and nitrogen-centred profiles simultaneously, hence accounting for uncertainties in the background level.
Uncertainties on the parameters were estimated using a jackknife resampling technique, that successively removed one of the lines from the combinations. This allowed us to estimate uncertainties due to potential unidentified lines in the selected ranges or local over/under-subtraction of the continuum. We measure $^{12}$C/$^{13}$C~=~$147^{+87}_{-40}$ and $^{14}$N/$^{15}$N~=~$343^{+454}_{-124}$. The values measured using this technique were fully consistent with those obtained through a fit using the python MCMC $emcee$ package that included parameters for the background level of the $^{13}$CN-centred and C$^{15}$N-centred profiles respectively. The uncertainties obtained through the jackknife resampling are also of the same order of magnitude as the 3-sigmas uncertainties coming from the MCMC fit.

\begin{longtable}{llcc}
\caption{ Nitrogen isotope ratios plotted in Figure~\ref{Nisotopes}. }
\label{tab:n_ratios}
\\
\hline
\textbf{Species} & \textbf{Comet} & \textbf{$^{14}\rm{N}/^{15}\rm{N}$} & \textbf{Reference} \\ 
\hline 
\endfirsthead

\hline 
\textbf{Species} & \textbf{Comet} & \textbf{$^{14}\rm{N}/^{15}\rm{N}$} & \textbf{Reference} \\ 
\hline 
\endhead

\hline 
\endfoot

\hline
\endlastfoot

\centering

HCN & C/1995 O1 (Hale-Bopp) & $205~{\pm}~70$ & \citep{Bockelee-Morvan2008} \\
 & C/2014 Q2 (Lovejoy) & $145~{\pm}~12$ & \citep{Biver2016} \\
& 17P/Holmes & $139~{\pm}~26$ & \citep{Biver2016} \\
 & 46P/Wirtanen & $68~{\pm}~27$ & \citep{Cordiner2024} \\

\hline

CN & C/1995 O1 (Hale-Bopp)& $130~{\pm}~40$ & \citep{Manfroid2005}, revised in \citep{Manfroid2009} \\
& & $150~{\pm}~30$ & \citep{Manfroid2005}, revised in \citep{Manfroid2009} \\
& & $135~{\pm}~40$ & \citep{Manfroid2005}, revised in \citep{Manfroid2009} \\
& C/1999 S4 (LINEAR) & $150~{\pm}~40$ & \citep{Hutsemekers2005}, revised in \citep{Manfroid2009} \\
& C/2000 WM1 (LINEAR) & $150~{\pm}~30$ & \citep{Arpigny2003}, revised in \citep{Manfroid2009} \\
& C/2001 Q4 (NEAT) & $130~{\pm}~40$ & \citep{Manfroid2005} \\
& C/2002 T7 (LINEAR) & $160~{\pm}~25$& \citep{Manfroid2009} \\
& C/2002 V1 (NEAT) & $160~{\pm}~35$ & \citep{Manfroid2009} \\
 & C/2002 X5 (Kudo-Fujikawa) & $130~{\pm}~20$ & \citep{Manfroid2009} \\
& C/2002 Y1 (Juels-Holvorcem) & $150~{\pm}~35$ & \citep{Manfroid2009} \\
& C/2003 K4 (LINEAR) & $150~{\pm}~35$ & \citep{Manfroid2005}, revised in \citep{Manfroid2009} \\
&  & $145~{\pm}~25$ & \citep{Manfroid2005}, revised in \citep{Manfroid2009} \\
& C/2006 M4 (SWAN) & $145~{\pm}~50$ & \citep{Manfroid2009} \\
& C/2007 N3 (Lulin) & $150~{\pm}~50$ & \citep{Manfroid2009} \\
& C/2015 ER61	(PanSTARRS) & $130~{\pm}~15$ &	\citep{Yang2018} \\
& 8P/Tuttle	& $150~{\pm}~30$ & \citep{Bockelee-Morvan2008_Tuttle} \\
& 9P/Tempel 1 & $110~{\pm}~20$ & \citep{Jehin2006ApJ} \\
& & $45~{\pm}~20$ & \citep{Jehin2006ApJ}  \\
& 17P/Holmes & $165~{\pm}~35$ &  \citep{Bockelee-Morvan2008} \\
& & $150~{\pm}~50$ &  \citep{Manfroid2009} \\
& 21P/Giacobini-Zinner & $145~{\pm}~10$ & \citep{Moulane2020} \\
& 46P/Wirtanen & $150~{\pm}~30$ &  \citep{Moulane2023}\\
& 73P-B/Schwassmann-Wachmann 3 & $210~{\pm}~50$ & \citep{Jehin2008} \\
& 73P-C/Schwassmann-Wachmann 3 & $220~{\pm}~40$ & \citep{Jehin2008} \\
& 88P/Howell & $140~{\pm}~20$ & \citep{Hutsemekers2005} \\
& 103P/Hartley 2 & $155~{\pm}~25$ & \citep{Jehin2011} \\
& 122P/de Vico & $145~{\pm}~20$ & \citep{Jehin2004} \\
& 153P/Ikeya-Zhang & $140~{\pm}~50$ & \citep{Jehin2004} \\
\hline

NH$_2$ & C/2000 WM1 (LINEAR) & $132~{\pm}~31$ & \citep{Shinnaka2016}\\
  & C/2001 Q4 (NEAT) & $145~{\pm}~35$ & \citep{Shinnaka2016}\\
  &   & $126~{\pm}~30$ & \citep{Shinnaka2016}\\
  &  C/2002 T7 (LINEAR) & $138~{\pm}~25$ & \citep{Shinnaka2016}\\
  &  C/2002 V1 (NEAT)  & $132~{\pm}~32$ & \citep{Shinnaka2016}\\
  &  C/2002 X5 (Kudo-Fujikawa)  & $137~{\pm}~27$ & \citep{Shinnaka2016}\\
  &  C/2002 Y1 (Juels Holvorcem)  & $132~{\pm}~33$ & \citep{Shinnaka2016}\\
  &  C/2003 K4 (LINEAR)  & $153~{\pm}~63$ & \citep{Shinnaka2016}\\
  &  C/2009 P1 (Garradd)  & $154~{\pm}~41$ & \citep{Shinnaka2016}\\
  &     & $132~{\pm}~61$ & \citep{Shinnaka2016}\\
  &  C/2012 F6 (Lemmon)  & $155~{\pm}~28$ & \citep{Shinnaka2016}\\
    &    & $152~{\pm}~72$ & \citep{Biver2016}\\
  &  C/2012 S1 (ISON)  & $139~{\pm}~38$ & \citep{Shinnaka2014}\\
  &  C/2013 R1 (Lovejoy)  & $126~{\pm}~28$ & \citep{Shinnaka2016}\\
    &   & $100~{\pm}~50$ & \citep{Rousselot2012}\\
    &  C/2013 US10 (Catalina)  & $149~{\pm}~32$ & \citep{Shinnaka2016}\\
  &  C/2014 E2 (Jacques)  & $126~{\pm}~55$ & \citep{Shinnaka2016}\\
  &  C/2014 Q2 (Lovejoy)  & $126~{\pm}~25$ & \citep{Shinnaka2016b}\\
  & & $145~{\pm}~12$ & \citep{Biver2016}\\
  & C/2015 ER61 (PanSTARRS) & $140~{\pm}~28$ & \citep{Yang2018}\\
  
\hline
\end{longtable}

\begin{longtable}{llcc}

\caption{Carbon isotope ratios plotted in Figure \ref{Cisotopes}.}
\label{tab:c_ratios}
\\

\textbf{Species} & \textbf{Comet} &\textbf{$^{12}\rm{C}/^{13}\rm{C}$} & \textbf{Reference}\\ 
\hline 
\endfirsthead

\textbf{Species} & \textbf{Comet} & \textbf{$^{12}\rm{C}/^{13}\rm{C}$} & \textbf{Reference}\\ 
\hline 
\hline 
\endhead

\endfoot

\hline \hline
\endlastfoot

\centering

C$_2$ & C/1963 A1 (Ikeya) & $70~{\pm}~15$ & \citep{Stawikoski1964} \\
 & C/1969 T1 (Tago-Sato-Kosaka) & $100~{\pm}~20$ & \citep{Owen1973} \\
  & C/1973 E1 (Kohoutek) & $115^{+30}_{-20}$ & \citep{Danks1974} \\
  & & $135^{+65}_{-45}$ & \citep{Danks1974} \\
  & C/1975 N1 (Kobayashi-Berger-Milon) & $100^{+20}_{-30}$ & \citep{Vanysek1977} as reported in \citep{Wyckoff2000} \\ 
 & C/1975 V1 (West) & $60~{\pm}~15$ &  \citep{Lambert1983}\\
 & C/2001 Q4 (NEAT) &	$80~{\pm}~20$ &	 \citep{Rousselot2012}\\
 & C/2002 T7 (LINEAR) &	$85~{\pm}~20$ &	 \citep{Rousselot2012}\\
 & C/2012 S1 (ISON) &	$94~{\pm}~33$ &	 \citep{Shinnaka2014} as reported in \citep{Biver2024}\\
\hline
CN & C/1989 Q1 (Okazaki-Levy-Rudenko)	& $93~{\pm}~20$ &  \citep{Wyckoff2000}\\
& C/1989 X1 (Austin) & $85~{\pm}~20$ &  \citep{Wyckoff2000}\\
& C/1990 K1 (Levy) & $90~{\pm}~10$ &  \citep{Wyckoff2000}\\
& C/1995 O1 (Hale-Bopp)& $90~{\pm}~15$ & \citep{Lis1997_halebopp}\\
& & $90~{\pm}~20$ & \citep{Manfroid2005}, revised in \citep{Manfroid2009} \\
& & $100~{\pm}~30$ & \citep{Manfroid2005}, revised in \citep{Manfroid2009} \\
& & $80~{\pm}~25$ & \citep{Manfroid2005}, revised in \citep{Manfroid2009} \\
& C/1999 S4 (LINEAR) & $90~{\pm}~30$ & \citep{Hutsemekers2005}, revised in \citep{Manfroid2009} \\
& C/2000 WM1 (LINEAR) & $100~{\pm}~20$ & \citep{Arpigny2003}, revised in \citep{Manfroid2009} \\
& C/2001 Q4 (NEAT) & $90~{\pm}~15$ & \citep{Manfroid2005} \\
&  & $70~{\pm}~30$ & \citep{Manfroid2005} \\
& C/2002 T7 (LINEAR) & $85~{\pm}~20$& \citep{Manfroid2009} \\
& C/2002 V1 (NEAT) & $100~{\pm}~20$ & \citep{Manfroid2009} \\
 & C/2002 X5 (Kudo-Fujikawa) & $90~{\pm}~20$ & \citep{Manfroid2009} \\
& C/2002 Y1 (Juels-Holvorcem) & $90~{\pm}~20$ & \citep{Manfroid2009} \\
& C/2003 K4 (LINEAR) & $90~{\pm}~20$ & \citep{Manfroid2005}, revised in \citep{Manfroid2009} \\
&  & $80~{\pm}~20$ & \citep{Manfroid2005}, revised in \citep{Manfroid2009} \\
& C/2006 M4 (SWAN) & $95~{\pm}~25$ & \citep{Manfroid2009} \\
& C/2007 N3 (Lulin) & $105~{\pm}~40$ & \citep{Manfroid2009} \\
& C/2012 F6 (Lemmon) & $95~{\pm}~25$ & \citep{Biver2024}\\
& C/2015 ER61	(PanSTARRS) & $100~{\pm}~15$ &	\citep{Yang2018} \\
& 1P/Halley & $65~{\pm}~9$ & \citep{Wyckoff1989}\\
& & $89~{\pm}~17$ & \citep{Jaworski1991}\\
& & $95~{\pm}~12$ & \citep{Kleine1995}\\
& 8P/Tuttle	& $90~{\pm}~20$ & \citep{Bockelee-Morvan2008_Tuttle} \\
& 9P/Tempel 1 & $95~{\pm}~15$ & \citep{Jehin2006ApJ} \\
& & $110~{\pm}~20$ & \citep{Jehin2006ApJ} as reported in \citep{Manfroid2009} \\
& 17P/Holmes & $90~{\pm}~20$ &  \citep{Bockelee-Morvan2008} \\
& & $90~{\pm}~20$ &  \citep{Manfroid2009} \\
& 21P/Giacobini-Zinner & $100~{\pm}~10$ & \citep{Moulane2020} \\
& 46P/Wirtanen & $100~{\pm}~20$ &  \citep{Moulane2023}\\
& 73P-B/Schwassmann-Wachmann 3 & $100~{\pm}~30$ & \citep{Jehin2008} \\
& 73P-C/Schwassmann-Wachmann 3 & $100~{\pm}~20$ & \citep{Jehin2008} \\
& 88P/Howell & $90~{\pm}~10$ & \citep{Hutsemekers2005} \\
& 103P/Hartley 2 & $95~{\pm}~15$ & \citep{Jehin2011} \\
& 122P/de Vico & $90~{\pm}~10$ & \citep{Jehin2004} \\
& 153P/Ikeya-Zhang & $80~{\pm}~30$ & \citep{Jehin2004} \\

& 3I/ATLAS & $147^{+87}_{-40}$ & this work \\
\hline

HCN & C/1995 O1 (Hale-Bopp) & $111~{\pm}~12$ & \citep{Jewitt1997} \\
& & $94~{\pm}~8$ & \citep{Bockelee-Morvan2008} \\
& & $109~{\pm}~22$ & \citep{Ziurys1999} \\
& C/1996 B2 (Hyakutake) & $34~{\pm}~12$ & \citep{Lis1997_hyakutake}\\
& C/2012 F6 (Lemmon) & $124~{\pm}~64$ & \citep{Biver2016} \\
& C/2012 S1 (ISON) & $88~{\pm}~18$ & \citep{Cordiner2019} \\
& C/2014 Q2 (Lovejoy) & $109~{\pm}~14$ & \citep{Biver2016} \\
& C/2022 E3 (ZTF) & $135~{\pm}~75$ & \citep{Biver2024} \\
& 17P/Holmes & $114~{\pm}~26$ & \citep{Bockelee-Morvan2008} \\
& 46P/Wirtanen & $90~{\pm}~28$ & \citep{Cordiner2024}\\
\hline

CO+ & C/2016 R2 (PanSTARRS) & $73~{\pm}~20$ &  \citep{Rousselot2024}\\
\hline

CO & 67P/Churyumov-Gerasimenko & $86~{\pm}~9$ & \citep{Rubin2017} \\
\hline

CO$_2$ & 67P/Churyumov-Gerasimenko & $84~{\pm}~4$ & \citep{Hassig2017} \\
\hline

CH$_4$ & 67P/Churyumov-Gerasimenko & $88~{\pm}~10$ & \citep{Muller2022} \\
\hline

C$_2$H$_6$ & 67P/Churyumov-Gerasimenko & $93~{\pm}~10$ & \citep{Muller2022} \\
 &  & $105~{\pm}~10$ & \citep{Altwegg2020}  \\
\hline

C$_3$H$_8$ & 67P/Churyumov-Gerasimenko & $87~{\pm}~9$ & \citep{Muller2022} \\
\hline

C$_4$H$_{10}$ & 67P/Churyumov-Gerasimenko & $96~{\pm}~14$ & \citep{Muller2022} \\
\hline

H$_2$CO & 67P/Churyumov-Gerasimenko & $40~{\pm}~14$ & \citep{Altwegg2020} \\
\hline

CH$_3$OH & 67P/Churyumov-Gerasimenko & $91~{\pm}~10$ & \citep{Altwegg2020}  \\

\hline
\end{longtable}

\backmatter


\section*{Acknowledgements}
The authors wish to acknowledge the exceptional level of support provided by support astronomers, Telescope and Instrument Operators, and User Support Department at Paranal Observatory. 
CO acknowledges the support of the Royal Society under grant URF\textbackslash R1\textbackslash211429.
MAC was supported by NASA’s Planetary Science Division Internal Scientist Funding Program through the Fundamental Laboratory Research work package (FLaRe).
The views expressed in this article are those of the authors and do not reflect the official policy or position of the U.S. Naval Academy, Department of the Navy, the Department of Defense, or the U.S. Government.
R.C.D. acknowledges support from grant \#361233 awarded by the Research Council of Finland to M. Granvik. J.M. is honorary Research Director at the F.R.S-FNRS. D.H. and E.J. are Research Directors at the F.R.S.-FNRS. 
For the purpose of open access, the authors have applied a Creative Commons Attribution (CC BY) licence to any Author Accepted Manuscript version arising from this submission.


\section*{Data availability}
All data used for this work will be available on the European Southern Observatory archive after a proprietary period of one year from the time of observations. They can be accessed searching for the programme number of the observations: 15.28NL.003.

\section*{Author contribution}
CO led the data acquisition and the writing of the manuscript. DH led the data reduction. JM led the data analysis. 
MMK and FL contributed figures and participated to the discussion of the results.
EJ, KA, DB, VVG, MC, RCD, LFF, ML, BPM, and SC contributed to the manuscript and discussion of the results.


\begin{thebibliography}{10}
\expandafter\ifx\csname url\endcsname\relax
  \def\url#1{\burl{#1}}\fi
\expandafter\ifx\csname urlprefix\endcsname\relax\def\urlprefix{URL }\fi
\providecommand{\bibinfo}[2]{#2}
\providecommand{\eprint}[2][]{\url{#2}}
\providecommand{\doi}[1]{\url{https://doi.org/#1}}
\bibcommenthead

\bibitem{Williams2017}
\bibinfo{author}{{Williams}, G.~V.} \emph{et~al.}
\newblock \bibinfo{title}{{Comet C/2017 u1 (panstarrs)}}.
\newblock \emph{\bibinfo{journal}{Minor Planet Electronic Circulars}} \textbf{\bibinfo{volume}{2017-U181}} (\bibinfo{year}{2017}).

\bibitem{Denneau2025}
\bibinfo{author}{{Denneau}, L.} \emph{et~al.}
\newblock \bibinfo{title}{{3I/ATLAS = C/2025 n1 (atlas)}}.
\newblock \emph{\bibinfo{journal}{Minor Planet Electronic Circulars}} \textbf{\bibinfo{volume}{2025-N12}} (\bibinfo{year}{2025}).

\bibitem{Jewitt2025a}
\bibinfo{author}{{Jewitt}, D.}, \bibinfo{author}{{Hui}, M.-T.}, \bibinfo{author}{{Mutchler}, M.}, \bibinfo{author}{{Kim}, Y.} \& \bibinfo{author}{{Agarwal}, J.}
\newblock \bibinfo{title}{{Hubble Space Telescope Observations of the Interstellar Interloper 3I/ATLAS}}.
\newblock \emph{\bibinfo{journal}{\apjl}} \textbf{\bibinfo{volume}{990}}, \bibinfo{pages}{L2} (\bibinfo{year}{2025}).

\bibitem{Alarcon2025}
\bibinfo{author}{{Alarcon}, M.~R.} \emph{et~al.}
\newblock \bibinfo{title}{{Deep g'-band Imaging of Interstellar Comet 3I/ATLAS from the Two-meter Twin Telescope (TTT)}}.
\newblock \emph{\bibinfo{journal}{The Astronomer's Telegram}} \textbf{\bibinfo{volume}{17264}}, \bibinfo{pages}{1} (\bibinfo{year}{2025}).

\bibitem{Seligman2025}
\bibinfo{author}{{Seligman}, D.~Z.} \emph{et~al.}
\newblock \bibinfo{title}{{Discovery and Preliminary Characterization of a Third Interstellar Object: 3I/ATLAS}}.
\newblock \emph{\bibinfo{journal}{\apjl}} \textbf{\bibinfo{volume}{989}}, \bibinfo{pages}{L36} (\bibinfo{year}{2025}).

\bibitem{Cordiner2025}
\bibinfo{author}{{Cordiner}, M.~A.} \emph{et~al.}
\newblock \bibinfo{title}{{JWST Detection of a Carbon-dioxide-dominated Gas Coma Surrounding Interstellar Object 3I/ATLAS}}.
\newblock \emph{\bibinfo{journal}{\apjl}} \textbf{\bibinfo{volume}{991}}, \bibinfo{pages}{L43} (\bibinfo{year}{2025}).

\bibitem{Cordiner2026}
\bibinfo{author}{{Cordiner}, M.~A.} \emph{et~al.}
\newblock \bibinfo{title}{{JWST Spectroscopy of 3I/ATLAS: Isotopic Evidence for a Cold and Distant Origin}}.
\newblock \emph{Submitted}  (\bibinfo{year}{2026}).

\bibitem{Rahatgaonkar2025}
\bibinfo{author}{{Rahatgaonkar}, R.} \emph{et~al.}
\newblock \bibinfo{title}{{VLT observations of interstellar comet 3I/ATLAS II. From quiescence to glow: Dramatic rise of Ni I emission and incipient CN outgassing at large heliocentric distances}}.
\newblock \emph{\bibinfo{journal}{arXiv e-prints}} \bibinfo{pages}{arXiv:2508.18382} (\bibinfo{year}{2025}).

\bibitem{Hutsemekers2025}
\bibinfo{author}{{Hutsem{\'e}kers}, D.} \emph{et~al.}
\newblock \bibinfo{title}{{Pre-perihelion evolution of the NiI/FeI abundance ratio in the coma of the interstellar comet 3I/ATLAS: From extreme to normal}}.
\newblock \emph{\bibinfo{journal}{\aap}} \textbf{\bibinfo{volume}{706}}, \bibinfo{pages}{A43} (\bibinfo{year}{2026}).

\bibitem{Roth2025}
\bibinfo{author}{{Roth}, N.~X.} \emph{et~al.}
\newblock \bibinfo{title}{{CH$_3$OH and HCN in Interstellar Comet 3I/ATLAS Mapped with the ALMA Atacama Compact Array: Distinct Outgassing Behaviors and a Remarkably High CH$_3$OH/HCN Production Rate Ratio}}.
\newblock \emph{\bibinfo{journal}{arXiv e-prints}} \bibinfo{pages}{arXiv:2511.20845} (\bibinfo{year}{2025}).

\bibitem{Nomura2023}
\bibinfo{author}{{Nomura}, H.} \emph{et~al.}
\newblock \bibinfo{editor}{{Inutsuka}, S.}, \bibinfo{editor}{{Aikawa}, Y.}, \bibinfo{editor}{{Muto}, T.}, \bibinfo{editor}{{Tomida}, K.} \& \bibinfo{editor}{{Tamura}, M.} (eds) \emph{\bibinfo{title}{{The Isotopic Links from Planet Forming Regions to the Solar System}}}.
\newblock (eds \bibinfo{editor}{{Inutsuka}, S.}, \bibinfo{editor}{{Aikawa}, Y.}, \bibinfo{editor}{{Muto}, T.}, \bibinfo{editor}{{Tomida}, K.} \& \bibinfo{editor}{{Tamura}, M.}) \emph{\bibinfo{booktitle}{Protostars and Planets VII}}, Vol. \bibinfo{volume}{534} of \emph{\bibinfo{series}{Astronomical Society of the Pacific Conference Series}}, \bibinfo{pages}{1075} (\bibinfo{year}{2023}).

\bibitem{Wirstrom2012}
\bibinfo{author}{{Wirstr{\"o}m}, E.~S.}, \bibinfo{author}{{Charnley}, S.~B.}, \bibinfo{author}{{Cordiner}, M.~A.} \& \bibinfo{author}{{Milam}, S.~N.}
\newblock \bibinfo{title}{{Isotopic Anomalies in Primitive Solar System Matter: Spin-state-dependent Fractionation of Nitrogen and Deuterium in Interstellar Clouds}}.
\newblock \emph{\bibinfo{journal}{\apjl}} \textbf{\bibinfo{volume}{757}}, \bibinfo{pages}{L11} (\bibinfo{year}{2012}).

\bibitem{Wirstrom2018}
\bibinfo{author}{{Wirstr{\"o}m}, E.~S.} \& \bibinfo{author}{{Charnley}, S.~B.}
\newblock \bibinfo{title}{{Revised models of interstellar nitrogen isotopic fractionation}}.
\newblock \emph{\bibinfo{journal}{\mnras}} \textbf{\bibinfo{volume}{474}}, \bibinfo{pages}{3720--3726} (\bibinfo{year}{2018}).

\bibitem{Hily-Blant2019}
\bibinfo{author}{{Hily-Blant}, P.}, \bibinfo{author}{{Magalhaes de Souza}, V.}, \bibinfo{author}{{Kastner}, J.} \& \bibinfo{author}{{Forveille}, T.}
\newblock \bibinfo{title}{{Multiple nitrogen reservoirs in a protoplanetary disk at the epoch of comet and giant planet formation}}.
\newblock \emph{\bibinfo{journal}{\aap}} \textbf{\bibinfo{volume}{632}}, \bibinfo{pages}{L12} (\bibinfo{year}{2019}).

\bibitem{Rampinelli2025}
\bibinfo{author}{{Rampinelli}, L.} \emph{et~al.}
\newblock \bibinfo{title}{{Radial variations in the nitrogen, carbon, and hydrogen fractionation in the PDS 70 planet-hosting disk}}.
\newblock \emph{\bibinfo{journal}{\aap}} \textbf{\bibinfo{volume}{698}}, \bibinfo{pages}{A115} (\bibinfo{year}{2025}).

\bibitem{Manfroid2009}
\bibinfo{author}{{Manfroid}, J.} \emph{et~al.}
\newblock \bibinfo{title}{{The CN isotopic ratios in comets}}.
\newblock \emph{\bibinfo{journal}{\aap}} \textbf{\bibinfo{volume}{503}}, \bibinfo{pages}{613--624} (\bibinfo{year}{2009}).

\bibitem{Biver2024}
\bibinfo{author}{{Biver}, N.} \emph{et~al.}
\newblock \bibinfo{title}{{Chemical composition of comets C/2021 A1 (Leonard) and C/2022 E3 (ZTF) from radio spectroscopy and the abundance of HCOOH and HNCO in comets}}.
\newblock \emph{\bibinfo{journal}{\aap}} \textbf{\bibinfo{volume}{690}}, \bibinfo{pages}{A271} (\bibinfo{year}{2024}).

\bibitem{Altwegg2019}
\bibinfo{author}{{Altwegg}, K.}, \bibinfo{author}{{Balsiger}, H.} \& \bibinfo{author}{{Fuselier}, S.~A.}
\newblock \bibinfo{title}{{Cometary Chemistry and the Origin of Icy Solar System Bodies: The View After Rosetta}}.
\newblock \emph{\bibinfo{journal}{\araa}} \textbf{\bibinfo{volume}{57}}, \bibinfo{pages}{113--155} (\bibinfo{year}{2019}).

\bibitem{Cordiner2024}
\bibinfo{author}{{Cordiner}, M.~A.} \emph{et~al.}
\newblock \bibinfo{title}{{Evidence for Surprising Heavy Nitrogen Isotopic Enrichment in Comet 46P/Wirtanen's Hydrogen Cyanide}}.
\newblock \emph{\bibinfo{journal}{\psj}} \textbf{\bibinfo{volume}{5}}, \bibinfo{pages}{221} (\bibinfo{year}{2024}).

\bibitem{Moulane2023}
\bibinfo{author}{{Moulane}, Y.} \emph{et~al.}
\newblock \bibinfo{title}{{Activity and composition of the hyperactive comet 46P/Wirtanen during its close approach in 2018}}.
\newblock \emph{\bibinfo{journal}{\aap}} \textbf{\bibinfo{volume}{670}}, \bibinfo{pages}{A159} (\bibinfo{year}{2023}).

\bibitem{Langer1984}
\bibinfo{author}{{Langer}, W.~D.}, \bibinfo{author}{{Graedel}, T.~E.}, \bibinfo{author}{{Frerking}, M.~A.} \& \bibinfo{author}{{Armentrout}, P.~B.}
\newblock \bibinfo{title}{{Carbon and oxygen isotope fractionation in dense interstellar clouds.}}
\newblock \emph{\bibinfo{journal}{\apj}} \textbf{\bibinfo{volume}{277}}, \bibinfo{pages}{581--604} (\bibinfo{year}{1984}).

\bibitem{Rubin2017}
\bibinfo{author}{{Rubin}, M.} \emph{et~al.}
\newblock \bibinfo{title}{{Evidence for depletion of heavy silicon isotopes at comet 67P/Churyumov-Gerasimenko}}.
\newblock \emph{\bibinfo{journal}{\aap}} \textbf{\bibinfo{volume}{601}}, \bibinfo{pages}{A123} (\bibinfo{year}{2017}).

\bibitem{Hassig2017}
\bibinfo{author}{{H{\"a}ssig}, M.} \emph{et~al.}
\newblock \bibinfo{title}{{Isotopic composition of CO$_{2}$ in the coma of 67P/Churyumov-Gerasimenko measured with ROSINA/DFMS}}.
\newblock \emph{\bibinfo{journal}{\aap}} \textbf{\bibinfo{volume}{605}}, \bibinfo{pages}{A50} (\bibinfo{year}{2017}).

\bibitem{Altwegg2020}
\bibinfo{author}{{Altwegg}, K.} \emph{et~al.}
\newblock \bibinfo{title}{{Molecule-dependent oxygen isotopic ratios in the coma of comet 67P/Churyumov-Gerasimenko}}.
\newblock \emph{\bibinfo{journal}{\mnras}} \textbf{\bibinfo{volume}{498}}, \bibinfo{pages}{5855--5862} (\bibinfo{year}{2020}).

\bibitem{Ritchey2015}
\bibinfo{author}{{Ritchey}, A.~M.}, \bibinfo{author}{{Federman}, S.~R.} \& \bibinfo{author}{{Lambert}, D.~L.}
\newblock \bibinfo{title}{{The C$^{14}$N/C$^{15}$N Ratio in Diffuse Molecular Clouds}}.
\newblock \emph{\bibinfo{journal}{\apjl}} \textbf{\bibinfo{volume}{804}}, \bibinfo{pages}{L3} (\bibinfo{year}{2015}).

\bibitem{Marty2011}
\bibinfo{author}{{Marty}, B.}, \bibinfo{author}{{Chaussidon}, M.}, \bibinfo{author}{{Wiens}, R.~C.}, \bibinfo{author}{{Jurewicz}, A.~J.~G.} \& \bibinfo{author}{{Burnett}, D.~S.}
\newblock \bibinfo{title}{{A $^{15}$N-Poor Isotopic Composition for the Solar System As Shown by Genesis Solar Wind Samples}}.
\newblock \emph{\bibinfo{journal}{Science}} \textbf{\bibinfo{volume}{332}}, \bibinfo{pages}{1533} (\bibinfo{year}{2011}).

\bibitem{Jensen2024}
\bibinfo{author}{{Jensen}, S.~S.} \emph{et~al.}
\newblock \bibinfo{title}{{Fractionation in young cores: Direct determinations of nitrogen and carbon fractionation in HCN}}.
\newblock \emph{\bibinfo{journal}{\aap}} \textbf{\bibinfo{volume}{685}}, \bibinfo{pages}{A149} (\bibinfo{year}{2024}).

\bibitem{Colzi2022}
\bibinfo{author}{{Colzi}, L.} \emph{et~al.}
\newblock \bibinfo{title}{{CHEMOUT: CHEMical complexity in star-forming regions of the OUTer Galaxy. III. Nitrogen isotopic ratios in the outer Galaxy}}.
\newblock \emph{\bibinfo{journal}{\aap}} \textbf{\bibinfo{volume}{667}}, \bibinfo{pages}{A151} (\bibinfo{year}{2022}).

\bibitem{Guzman2017}
\bibinfo{author}{{Guzm{\'a}n}, V.~V.}, \bibinfo{author}{{{\"O}berg}, K.~I.}, \bibinfo{author}{{Huang}, J.}, \bibinfo{author}{{Loomis}, R.} \& \bibinfo{author}{{Qi}, C.}
\newblock \bibinfo{title}{{Nitrogen Fractionation in Protoplanetary Disks from the H$^{13}$CN/HC$^{15}$N Ratio}}.
\newblock \emph{\bibinfo{journal}{\apj}} \textbf{\bibinfo{volume}{836}}, \bibinfo{pages}{30} (\bibinfo{year}{2017}).

\bibitem{Wilson1999}
\bibinfo{author}{{Wilson}, T.~L.}
\newblock \bibinfo{title}{{Isotopes in the interstellar medium and circumstellar envelopes}}.
\newblock \emph{\bibinfo{journal}{Reports on Progress in Physics}} \textbf{\bibinfo{volume}{62}}, \bibinfo{pages}{143--185} (\bibinfo{year}{1999}).

\bibitem{Hopkins2025}
\bibinfo{author}{{Hopkins}, M.~J.} \emph{et~al.}
\newblock \bibinfo{title}{{From a Different Star: 3I/ATLAS in the Context of the {\={O}}tautahi{\textendash}Oxford Interstellar Object Population Model}}.
\newblock \emph{\bibinfo{journal}{\apjl}} \textbf{\bibinfo{volume}{990}}, \bibinfo{pages}{L30} (\bibinfo{year}{2025}).

\bibitem{Taylor2025}
\bibinfo{author}{{Taylor}, A.~G.} \& \bibinfo{author}{{Seligman}, D.~Z.}
\newblock \bibinfo{title}{{The Kinematic Age of 3I/ATLAS and Its Implications for Early Planet Formation}}.
\newblock \emph{\bibinfo{journal}{\apjl}} \textbf{\bibinfo{volume}{990}}, \bibinfo{pages}{L14} (\bibinfo{year}{2025}).

\bibitem{Yan2023}
\bibinfo{author}{{Yan}, Y.~T.} \emph{et~al.}
\newblock \bibinfo{title}{{Direct measurements of carbon and sulfur isotope ratios in the Milky Way}}.
\newblock \emph{\bibinfo{journal}{\aap}} \textbf{\bibinfo{volume}{670}}, \bibinfo{pages}{A98} (\bibinfo{year}{2023}).

\bibitem{Milam2005}
\bibinfo{author}{{Milam}, S.~N.}, \bibinfo{author}{{Savage}, C.}, \bibinfo{author}{{Brewster}, M.~A.}, \bibinfo{author}{{Ziurys}, L.~M.} \& \bibinfo{author}{{Wyckoff}, S.}
\newblock \bibinfo{title}{{The $^{12}$C/$^{13}$C Isotope Gradient Derived from Millimeter Transitions of CN: The Case for Galactic Chemical Evolution}}.
\newblock \emph{\bibinfo{journal}{\apj}} \textbf{\bibinfo{volume}{634}}, \bibinfo{pages}{1126--1132} (\bibinfo{year}{2005}).

\bibitem{Timmes1995}
\bibinfo{author}{{Timmes}, F.~X.}, \bibinfo{author}{{Woosley}, S.~E.} \& \bibinfo{author}{{Weaver}, T.~A.}
\newblock \bibinfo{editor}{{Busso}, M.}, \bibinfo{editor}{{Raiteri}, C.~M.} \& \bibinfo{editor}{{Gallino}, R.} (eds) \emph{\bibinfo{title}{{Galactic Chemical Evolution: Neutrino-Process Contributions}}}.
\newblock (eds \bibinfo{editor}{{Busso}, M.}, \bibinfo{editor}{{Raiteri}, C.~M.} \& \bibinfo{editor}{{Gallino}, R.}) \emph{\bibinfo{booktitle}{Nuclei in the Cosmos III}}, Vol. \bibinfo{volume}{327} of \emph{\bibinfo{series}{American Institute of Physics Conference Series}}, \bibinfo{pages}{543} (\bibinfo{publisher}{AIP}, \bibinfo{year}{1995}).

\bibitem{Kobayashi2011}
\bibinfo{author}{{Kobayashi}, C.}, \bibinfo{author}{{Karakas}, A.~I.} \& \bibinfo{author}{{Umeda}, H.}
\newblock \bibinfo{title}{{The evolution of isotope ratios in the Milky Way Galaxy}}.
\newblock \emph{\bibinfo{journal}{\mnras}} \textbf{\bibinfo{volume}{414}}, \bibinfo{pages}{3231--3250} (\bibinfo{year}{2011}).

\bibitem{Lee2024}
\bibinfo{author}{{Lee}, S.}, \bibinfo{author}{{Nomura}, H.} \& \bibinfo{author}{{Furuya}, K.}
\newblock \bibinfo{title}{{Carbon Isotope Chemistry in Protoplanetary Disks: Effects of C/O Ratios}}.
\newblock \emph{\bibinfo{journal}{\apj}} \textbf{\bibinfo{volume}{969}}, \bibinfo{pages}{41} (\bibinfo{year}{2024}).

\bibitem{Bedell2018}
\bibinfo{author}{{Bedell}, M.} \emph{et~al.}
\newblock \bibinfo{title}{{The Chemical Homogeneity of Sun-like Stars in the Solar Neighborhood}}.
\newblock \emph{\bibinfo{journal}{\apj}} \textbf{\bibinfo{volume}{865}}, \bibinfo{pages}{68} (\bibinfo{year}{2018}).

\bibitem{Nissen2015}
\bibinfo{author}{{Nissen}, P.~E.}
\newblock \bibinfo{title}{{High-precision abundances of elements in solar twin stars. Trends with stellar age and elemental condensation temperature}}.
\newblock \emph{\bibinfo{journal}{\aap}} \textbf{\bibinfo{volume}{579}}, \bibinfo{pages}{A52} (\bibinfo{year}{2015}).

\bibitem{Maggiolo2026}
\bibinfo{author}{{Maggiolo}, R.}, \bibinfo{author}{{Dhooghe}, F.}, \bibinfo{author}{{Gronoff}, G.~P.}, \bibinfo{author}{{de Keyser}, J.} \& \bibinfo{author}{{Cessateur}, G.}
\newblock \bibinfo{title}{{Interstellar Comet 3I/ATLAS: Evidence for Galactic Cosmic-Ray Processing}}.
\newblock \emph{\bibinfo{journal}{\apjl}} \textbf{\bibinfo{volume}{996}}, \bibinfo{pages}{L34} (\bibinfo{year}{2026}).

\bibitem{Zhang2017}
\bibinfo{author}{{Zhang}, K.}, \bibinfo{author}{{Bergin}, E.~A.}, \bibinfo{author}{{Blake}, G.~A.}, \bibinfo{author}{{Cleeves}, L.~I.} \& \bibinfo{author}{{Schwarz}, K.~R.}
\newblock \bibinfo{title}{{Mass inventory of the giant-planet formation zone in a solar nebula analogue}}.
\newblock \emph{\bibinfo{journal}{Nature Astronomy}} \textbf{\bibinfo{volume}{1}}, \bibinfo{pages}{0130} (\bibinfo{year}{2017}).

\bibitem{Yoshida2022}
\bibinfo{author}{{Yoshida}, T.~C.}, \bibinfo{author}{{Nomura}, H.}, \bibinfo{author}{{Furuya}, K.}, \bibinfo{author}{{Tsukagoshi}, T.} \& \bibinfo{author}{{Lee}, S.}
\newblock \bibinfo{title}{{A New Method for Direct Measurement of Isotopologue Ratios in Protoplanetary Disks: A Case Study of the $^{12}$CO/$^{13}$CO Ratio in the TW Hya Disk}}.
\newblock \emph{\bibinfo{journal}{\apj}} \textbf{\bibinfo{volume}{932}}, \bibinfo{pages}{126} (\bibinfo{year}{2022}).

\bibitem{vanDokkum2001}
\bibinfo{author}{{van Dokkum}, P.~G.}
\newblock \bibinfo{title}{{Cosmic-Ray Rejection by Laplacian Edge Detection}}.
\newblock \emph{\bibinfo{journal}{\pasp}} \textbf{\bibinfo{volume}{113}}, \bibinfo{pages}{1420--1427} (\bibinfo{year}{2001}).

\bibitem{Bockelee-Morvan2008}
\bibinfo{author}{{Bockel{\'e}e-Morvan}, D.} \emph{et~al.}
\newblock \bibinfo{title}{{Large Excess of Heavy Nitrogen in Both Hydrogen Cyanide and Cyanogen from Comet 17P/Holmes}}.
\newblock \emph{\bibinfo{journal}{\apjl}} \textbf{\bibinfo{volume}{679}}, \bibinfo{pages}{L49} (\bibinfo{year}{2008}).

\bibitem{Biver2016}
\bibinfo{author}{{Biver}, N.} \emph{et~al.}
\newblock \bibinfo{title}{{Isotopic ratios of H, C, N, O, and S in comets C/2012 F6 (Lemmon) and C/2014 Q2 (Lovejoy)}}.
\newblock \emph{\bibinfo{journal}{\aap}} \textbf{\bibinfo{volume}{589}}, \bibinfo{pages}{A78} (\bibinfo{year}{2016}).

\bibitem{Manfroid2005}
\bibinfo{author}{{Manfroid}, J.} \emph{et~al.}
\newblock \bibinfo{title}{{Isotopic abundance of nitrogen and carbon in distant comets}}.
\newblock \emph{\bibinfo{journal}{\aap}} \textbf{\bibinfo{volume}{432}}, \bibinfo{pages}{L5--L8} (\bibinfo{year}{2005}).

\bibitem{Hutsemekers2005}
\bibinfo{author}{{Hutsem{\'e}kers}, D.} \emph{et~al.}
\newblock \bibinfo{title}{{Isotopic abundances of carbon and nitrogen in Jupiter-family and Oort Cloud comets}}.
\newblock \emph{\bibinfo{journal}{\aap}} \textbf{\bibinfo{volume}{440}}, \bibinfo{pages}{L21--L24} (\bibinfo{year}{2005}).

\bibitem{Arpigny2003}
\bibinfo{author}{{Arpigny}, C.} \emph{et~al.}
\newblock \bibinfo{title}{{Anomalous Nitrogen Isotope Ratio in Comets}}.
\newblock \emph{\bibinfo{journal}{Science}} \textbf{\bibinfo{volume}{301}}, \bibinfo{pages}{1522--1525} (\bibinfo{year}{2003}).

\bibitem{Yang2018}
\bibinfo{author}{{Yang}, B.} \emph{et~al.}
\newblock \bibinfo{title}{{Isotopic ratios in outbursting comet C/2015 ER61}}.
\newblock \emph{\bibinfo{journal}{\aap}} \textbf{\bibinfo{volume}{609}}, \bibinfo{pages}{L4} (\bibinfo{year}{2018}).

\bibitem{Bockelee-Morvan2008_Tuttle}
\bibinfo{author}{{Bockel{\'e}e-Morvan}, D.} \emph{et~al.}
\newblock \bibinfo{editor}{{LPI Editorial Board}} (ed.) \emph{\bibinfo{title}{{A Multi-Wavelength Simultaneous Study of the Composition of the Halley-Family Comet 8P/Tuttle at the VLT}}}.
\newblock (ed.\bibinfo{editor}{{LPI Editorial Board}}) \emph{\bibinfo{booktitle}{Asteroids, Comets, Meteors 2008}}, Vol. \bibinfo{volume}{1405} of \emph{\bibinfo{series}{LPI Contributions}}, \bibinfo{pages}{8190} (\bibinfo{year}{2008}).

\bibitem{Jehin2006ApJ}
\bibinfo{author}{{Jehin}, E.} \emph{et~al.}
\newblock \bibinfo{title}{{Deep Impact: High-Resolution Optical Spectroscopy with the ESO VLT and the Keck I Telescope}}.
\newblock \emph{\bibinfo{journal}{\apjl}} \textbf{\bibinfo{volume}{641}}, \bibinfo{pages}{L145--L148} (\bibinfo{year}{2006}).

\bibitem{Moulane2020}
\bibinfo{author}{{Moulane}, Y.} \emph{et~al.}
\newblock \bibinfo{title}{{Photometry and high-resolution spectroscopy of comet 21P/Giacobini-Zinner during its 2018 apparition}}.
\newblock \emph{\bibinfo{journal}{\aap}} \textbf{\bibinfo{volume}{640}}, \bibinfo{pages}{A54} (\bibinfo{year}{2020}).

\bibitem{Jehin2008}
\bibinfo{author}{{Jehin}, E.} \emph{et~al.}
\newblock \bibinfo{editor}{{LPI Editorial Board}} (ed.) \emph{\bibinfo{title}{{Optical Spectroscopy of the B and C Fragments of Comet 73P/Schwassmann-Wachmann 3 at the ESO VLT}}}.
\newblock (ed.\bibinfo{editor}{{LPI Editorial Board}}) \emph{\bibinfo{booktitle}{Asteroids, Comets, Meteors 2008}}, Vol. \bibinfo{volume}{1405} of \emph{\bibinfo{series}{LPI Contributions}}, \bibinfo{pages}{8319} (\bibinfo{year}{2008}).

\bibitem{Jehin2011}
\bibinfo{author}{{Jehin}, E.} \emph{et~al.}
\newblock \emph{\bibinfo{title}{{A Multi-wavelength study with the ESO VLT of comet 103P/Hartley2 at the time of the EPOXI encounter}}}, Vol. \bibinfo{volume}{2011}, \bibinfo{pages}{1463} (\bibinfo{year}{2011}).

\bibitem{Jehin2004}
\bibinfo{author}{{Jehin}, E.} \emph{et~al.}
\newblock \bibinfo{title}{{The Anomalous $^{14}$N/$^{15}$N Ratio in Comets 122P/1995 S1 (de Vico) and 153P/2002 C1 (Ikeya-Zhang)}}.
\newblock \emph{\bibinfo{journal}{\apjl}} \textbf{\bibinfo{volume}{613}}, \bibinfo{pages}{L161--L164} (\bibinfo{year}{2004}).

\bibitem{Shinnaka2016}
\bibinfo{author}{{Shinnaka}, Y.} \emph{et~al.}
\newblock \bibinfo{title}{{Nitrogen isotopic ratios of NH$_{2}$ in comets: implication for $^{15}$N-fractionation in cometary ammonia}}.
\newblock \emph{\bibinfo{journal}{\mnras}} \textbf{\bibinfo{volume}{462}}, \bibinfo{pages}{S195--S209} (\bibinfo{year}{2016}).

\bibitem{Shinnaka2014}
\bibinfo{author}{{Shinnaka}, Y.} \emph{et~al.}
\newblock \emph{\bibinfo{title}{{High-Dispersion Spectroscopic Observations of Comet C/2012 S1 (ISON) with the Subaru Telescope}}}, Vol.~\bibinfo{volume}{46} of \emph{\bibinfo{series}{AAS/Division for Planetary Sciences Meeting Abstracts}}, \bibinfo{pages}{209.14} (\bibinfo{year}{2014}).

\bibitem{Rousselot2012}
\bibinfo{author}{{Rousselot}, P.}, \bibinfo{author}{{Jehin}, E.}, \bibinfo{author}{{Manfroid}, J.} \& \bibinfo{author}{{Hutsem{\'e}kers}, D.}
\newblock \bibinfo{title}{{The $^{12}$C$_{2}$/$^{12}$C$^{13}$C isotopic ratio in comets C/2001 Q4 (NEAT) and C/2002 T7 (LINEAR)}}.
\newblock \emph{\bibinfo{journal}{\aap}} \textbf{\bibinfo{volume}{545}}, \bibinfo{pages}{A24} (\bibinfo{year}{2012}).

\bibitem{Shinnaka2016b}
\bibinfo{author}{{Shinnaka}, Y.} \& \bibinfo{author}{{Kawakita}, H.}
\newblock \bibinfo{title}{{Nitrogen Isotopic Ratio of Cometary Ammonia from High-resolution Optical Spectroscopic Observations of C/2014 Q2 (Lovejoy)}}.
\newblock \emph{\bibinfo{journal}{\aj}} \textbf{\bibinfo{volume}{152}}, \bibinfo{pages}{145} (\bibinfo{year}{2016}).

\bibitem{Stawikoski1964}
\bibinfo{author}{{Stawikowski}, A.} \& \bibinfo{author}{{Greenstein}, J.~L.}
\newblock \bibinfo{title}{{The Isotope Ratio C\^\{12\}/C\^\{13\} in a Comet.}}
\newblock \emph{\bibinfo{journal}{\apj}} \textbf{\bibinfo{volume}{140}}, \bibinfo{pages}{1280} (\bibinfo{year}{1964}).

\bibitem{Owen1973}
\bibinfo{author}{{Owen}, T.}
\newblock \bibinfo{title}{{The Isotope Ratio 12C/13C in Comet Tago-Sato (1969g)}}.
\newblock \emph{\bibinfo{journal}{\apj}} \textbf{\bibinfo{volume}{184}}, \bibinfo{pages}{33--44} (\bibinfo{year}{1973}).

\bibitem{Danks1974}
\bibinfo{author}{{Danks}, A.~C.}, \bibinfo{author}{{Lambert}, D.~L.} \& \bibinfo{author}{{Arpigny}, C.}
\newblock \bibinfo{title}{{The $^{12}$C/$^{13}$C ratio in comet Kohoutek (1973f).}}
\newblock \emph{\bibinfo{journal}{\apj}} \textbf{\bibinfo{volume}{194}}, \bibinfo{pages}{745--751} (\bibinfo{year}{1974}).

\bibitem{Vanysek1977}
\bibinfo{author}{{Vanysek}, V.}
\newblock \bibinfo{editor}{{Delsemme}, A.~H.} (ed.) \emph{\bibinfo{title}{{Carbon isotope ratio in comets and interstellar matter.}}}
\newblock (ed.\bibinfo{editor}{{Delsemme}, A.~H.}) \emph{\bibinfo{booktitle}{IAU Colloquium 39: Comets, Asteroids, Meteorites: Interrelations, Evolution and Origins}}, \bibinfo{pages}{499--503} (\bibinfo{year}{1977}).

\bibitem{Wyckoff2000}
\bibinfo{author}{{Wyckoff}, S.}, \bibinfo{author}{{Kleine}, M.}, \bibinfo{author}{{Peterson}, B.~A.}, \bibinfo{author}{{Wehinger}, P.~A.} \& \bibinfo{author}{{Ziurys}, L.~M.}
\newblock \bibinfo{title}{{Carbon Isotope Abundances in Comets}}.
\newblock \emph{\bibinfo{journal}{\apj}} \textbf{\bibinfo{volume}{535}}, \bibinfo{pages}{991--999} (\bibinfo{year}{2000}).

\bibitem{Lambert1983}
\bibinfo{author}{{Lambert}, D.~L.} \& \bibinfo{author}{{Danks}, A.~C.}
\newblock \bibinfo{title}{{High-resolution spectra of C2 Swan bands from comet West 1976 VI}}.
\newblock \emph{\bibinfo{journal}{\apj}} \textbf{\bibinfo{volume}{268}}, \bibinfo{pages}{428--446} (\bibinfo{year}{1983}).

\bibitem{Lis1997_halebopp}
\bibinfo{author}{{Lis}, D.~C.} \emph{et~al.}
\newblock \bibinfo{title}{{Spectroscopic Observations of Comet C/1996 B2 (Hyakutake) with the Caltech Submillimeter Observatory}}.
\newblock \emph{\bibinfo{journal}{\icarus}} \textbf{\bibinfo{volume}{130}}, \bibinfo{pages}{355--372} (\bibinfo{year}{1997}).

\bibitem{Wyckoff1989}
\bibinfo{author}{{Wyckoff}, S.} \emph{et~al.}
\newblock \bibinfo{title}{{The 12C/ 13C Abundance Ratio in Comet Halley}}.
\newblock \emph{\bibinfo{journal}{\apj}} \textbf{\bibinfo{volume}{339}}, \bibinfo{pages}{488} (\bibinfo{year}{1989}).

\bibitem{Jaworski1991}
\bibinfo{author}{{Jaworski}, W.~A.} \& \bibinfo{author}{{Tatum}, J.~B.}
\newblock \bibinfo{title}{{Analysis of the Swings Effect and Greenstein Effect in Comet P/Halley}}.
\newblock \emph{\bibinfo{journal}{\apj}} \textbf{\bibinfo{volume}{377}}, \bibinfo{pages}{306} (\bibinfo{year}{1991}).

\bibitem{Kleine1995}
\bibinfo{author}{{Kleine}, M.}, \bibinfo{author}{{Wyckoff}, S.}, \bibinfo{author}{{Wehinger}, P.~A.} \& \bibinfo{author}{{Peterson}, B.~A.}
\newblock \bibinfo{title}{{The Carbon Isotope Abundance Ratio in Comet Halley}}.
\newblock \emph{\bibinfo{journal}{\apj}} \textbf{\bibinfo{volume}{439}}, \bibinfo{pages}{1021} (\bibinfo{year}{1995}).

\bibitem{Jewitt1997}
\bibinfo{author}{{Jewitt}, D.}, \bibinfo{author}{{Matthews}, H.~E.}, \bibinfo{author}{{Owen}, T.} \& \bibinfo{author}{{Meier}, R.}
\newblock \bibinfo{title}{{The 12C/13C, 14N/15N and 32S/ 34S Isotope Ratios in Comet Hale-Bopp (C/1995 O1).}}
\newblock \emph{\bibinfo{journal}{Science}} \textbf{\bibinfo{volume}{278}}, \bibinfo{pages}{90--93} (\bibinfo{year}{1997}).

\bibitem{Ziurys1999}
\bibinfo{author}{{Ziurys}, L.~M.} \emph{et~al.}
\newblock \bibinfo{title}{{Cyanide Chemistry in Comet Hale-Bopp (C/1995 O1)}}.
\newblock \emph{\bibinfo{journal}{\apjl}} \textbf{\bibinfo{volume}{527}}, \bibinfo{pages}{L67--L71} (\bibinfo{year}{1999}).

\bibitem{Lis1997_hyakutake}
\bibinfo{author}{{Lis}, D.~C.} \emph{et~al.}
\newblock \bibinfo{title}{{Spectroscopic Observations of Comet C/1996 B2 (Hyakutake) with the Caltech Submillimeter Observatory}}.
\newblock \emph{\bibinfo{journal}{\icarus}} \textbf{\bibinfo{volume}{130}}, \bibinfo{pages}{355--372} (\bibinfo{year}{1997}).

\bibitem{Cordiner2019}
\bibinfo{author}{{Cordiner}, M.~A.} \emph{et~al.}
\newblock \bibinfo{title}{{ALMA Autocorrelation Spectroscopy of Comets: The HCN/H$^{13}$CN Ratio in C/2012 S1 (ISON)}}.
\newblock \emph{\bibinfo{journal}{\apjl}} \textbf{\bibinfo{volume}{870}}, \bibinfo{pages}{L26} (\bibinfo{year}{2019}).

\bibitem{Rousselot2024}
\bibinfo{author}{{Rousselot}, P.} \emph{et~al.}
\newblock \bibinfo{title}{{$^{12}$CO$^{+}$ and $^{13}$CO$^{+}$ fluorescence models for measuring the $^{12}$C/$^{13}$C isotopic ratio in comets}}.
\newblock \emph{\bibinfo{journal}{\aap}} \textbf{\bibinfo{volume}{683}}, \bibinfo{pages}{A50} (\bibinfo{year}{2024}).

\bibitem{Muller2022}
\bibinfo{author}{{M{\"u}ller}, D.~R.} \emph{et~al.}
\newblock \bibinfo{title}{{High D/H ratios in water and alkanes in comet 67P/Churyumov-Gerasimenko measured with Rosetta/ROSINA DFMS}}.
\newblock \emph{\bibinfo{journal}{\aap}} \textbf{\bibinfo{volume}{662}}, \bibinfo{pages}{A69} (\bibinfo{year}{2022}).

\end{thebibliography}
\end{document}